\documentclass[fleqn,usenatbib]{mnras}
\usepackage{comment}

\usepackage{newtxtext,newtxmath}

\usepackage[T1]{fontenc}
\usepackage{ae,aecompl}

\usepackage{graphicx}	
\usepackage{amsmath}	
\usepackage{amssymb}	

\usepackage{xcolor}

\def\be{\begin{equation}}
\def\ee{\end{equation}}
\newcommand\code[1]{\textsc{\MakeLowercase{#1}}}

\newcommand\quotes[1]{``{#1}"}
\def\gsim{\lower.5ex\hbox{\gtsima}} 
\def\lsim{\lower.5ex\hbox{\ltsima}} 
\def\gtsima{$\; \buildrel > \over \sim \;$} 
\def\ltsima{$\; \buildrel < \over \sim \;$} \def\gsim{\lower.5ex\hbox{\gtsima}} 
\def\lsim{\lower.5ex\hbox{\ltsima}} 
\def\simgt{\lower.5ex\hbox{\gtsima}} 
\def\simlt{\lower.5ex\hbox{\ltsima}}

\def\msun{{\rm M}_{\odot}}

\def\mum{\mu {\rm m}}

\def\cc{\rm cm^{-3}}

\def\fdust{\xi_{d}}

\def\td{\tau_{sd}}

\def\S*{$\Sigma_{\rm SFR}$}
\def\Scii{$\Sigma_{\rm [CII]}$}
\def\Sciimax{$\Sigma_{\rm [CII]}^{\rm max}$}
\def\CII{\hbox{[C~$\scriptstyle\rm II $]~}}
\def\CIII{\hbox{C~$\scriptstyle\rm III $]~}}
\def\HH{\hbox{H$_2$}~} 
\def\HI{\hbox{H~$\scriptstyle\rm I\ $}} 
\def\HII{\hbox{H~$\scriptstyle\rm II\ $}} 
\def\CIion{\hbox{C~$\scriptstyle\rm I $~}}
\def\CIIion{\hbox{C~$\scriptstyle\rm II $~}}
\def\CIIIion{\hbox{C~$\scriptstyle\rm III $~}}
\def\CIVion{\hbox{C~$\scriptstyle\rm IV $~}}
\def\nhh{n_{\rm H2}}
\def\nhi{n_{\rm HI}}

\def\fhi{x_{\rm HI}}
\def\fhii{x_{\rm HII}}
\def\fd{f^*_{\rm diss}} 
\def\ks{\kappa_{\rm s}}

\definecolor{apcolor}{HTML}{b3003b}
\definecolor{afcolor}{HTML}{800080}
\definecolor{lvcolor}{HTML}{DF7401}
\definecolor{mdcolor}{HTML}{01abdf} 
\definecolor{cbcolor}{HTML}{ff0000}
\definecolor{sccolor}{HTML}{cc5500} 
\definecolor{sgcolor}{HTML}{00cc7a}

\defcitealias{carniani2018}{C18}

\title[\CII emission from galaxies]{A physical model for [CII] line emission from galaxies}

\author[Ferrara et al.]{A. Ferrara$^{1,6}$\thanks{E-mail: andrea.ferrara@sns.it},
L. Vallini$^{2,3}$,
A. Pallottini$^{1,4}$,
S. Gallerani$^{1}$,
S. Carniani$^{1}$,
M. Kohandel$^{1}$,\newauthor
D. Decataldo$^{1}$,
C. Behrens$^{1,5}$
\\
$^{1}$Scuola Normale Superiore, Piazza dei Cavalieri 7, 56126 Pisa, Italy\\
$^{2}$Leiden Observatory, Leiden University, PO Box 9500, 2300 RA Leiden, The Netherlands\\
$^{3}$Nordita, KTH Royal Institute of Technology and Stockholm University Roslagstullsbacken 23, SE-106 91 Stockholm, Sweden\\
$^{4}$Centro Fermi, Museo Storico della Fisica e Centro Studi e Ricerche \quotes{Enrico Fermi}, Piazza del Viminale 1, Roma, 00184, Italy\\
$^{5}$Institut f\"{u}r Astrophysik, Georg-August Universit\"{a}t G\"{o}ttingen, Friedrich-Hundt-Platz 1, 37077, G\"{o}ttingen, Germany\\
$^{6}$Kavli Institute for the Physics and Mathematics of the Universe (WPI), University of Tokyo, Kashiwa 277-8583, Japan\\
}

\date{Accepted XXX. Received YYY; in original form ZZZ}

\pubyear{2019}

\begin{document}
\label{firstpage}
\pagerange{\pageref{firstpage}--\pageref{lastpage}}
\maketitle

\begin{abstract}
A tight relation between the \CII 158$\mu$m line luminosity and star formation rate is measured in local galaxies. At high redshift ($z>5$), though, a much larger scatter is observed, with a considerable  (15-20\%) fraction of the outliers being [\ion{C}{ii}]-deficient. Moreover, the [CII] surface brightness (\Scii) of these sources is systematically lower than expected from the  local relation. To clarify the origin of such [\ion{C}{ii}]-deficiency we have developed an analytical model that fits local \CII data and has been validated against radiative transfer simulations performed with \code{cloudy}.  The model predicts an overall increase of \Scii~with \S*. However, for \S*$\simgt 1 M_\odot~{\rm yr}^{-1}~{\rm kpc}^{-2}$, \Scii~saturates. 
We conclude that underluminous \CII systems can result from a combination of three factors: (a) large upward deviations from the Kennicutt-Schmidt relation ($\ks\gg 1$), parameterized by the \quotes{burstiness} parameter $\ks$; (b) low metallicity; (c) low gas density, at least for the most extreme sources (e.g. CR7). Observations of \CII emission alone cannot break the degeneracy among the above three parameters; this requires additional information coming from other emission lines (e.g. [OIII]88$\mu$m, CIII]1909\AA, CO lines). Simple formulae are given to interpret available data for low and high-$z$ galaxies. 
\end{abstract}



\begin{table}
    \caption{List of main symbols in order of appearance}
    \label{tab:table1}
    \begin{tabular}{l|l|l} 
      \textbf{Symbol} & \textbf{Description} & \textbf{Section}\\
      \hline
      $n$               & Total gas density & 2\\
      $Z$               & Metallicity & 2\\
      $N_0$             & Total column density of the slab/galaxy & 2\\
      $F_i$             & Ionising photon flux  & 2.1\\
      $F_L$             & Radiation flux at the Lyman limit  & 2.1\\
      $\beta$           & Slope of the radiation spectrum & 2.1\\
      $U$               & Ionisation parameter & 2.1\\
      $\ell_S$          & Str\"omgren depth & 2.1\\
      $N_S$             & Str\"omgren column density & 2.1\\
      $\Gamma$          & photo-ionisation rate & 2.1\\
      $\tau_S$          & photo-ionisation optical depth to $\ell_S$ & 2.1\\
      $\tau$            & Total (gas + dust) UV optical depth & 2.1\\
      $x_{\rm HII}$     & Ionised hydrogen fraction & 2.1\\
      ${\cal D}$        & Dust-to-gas ratio & 2.1\\
      $\bar\sigma_d$    & Flux-weighted dust extinction cross-section  & 2.1\\
      $\tau_{sd}$       & Dust optical depth to $\ell_S$ & 2.1\\
      $x_{\rm HI}$      & Neutral hydrogen fraction & 2.1\\
      $N_d$             & Gas column density with dust $\tau_d=1$ & 2.1\\
      $N_i$             & Ionised gas column density & 2.1\\
      $N_{\rm HI}$      & Neutral hydrogen column density & 2.1\\
      $F_0$             & LW photon flux impinging on the cloud & 2.2\\
      ${\mathcal F}$    & Flux normalised to ${F_0}$ & 2.2 \\
      ${\cal R}$        & Rate of H$_2$ formation on grain surfaces & 2.2\\
      $\chi$            & Normalised H$_2$ photo-dissociation ratio & 2.2\\
      $w$               & Branching ratio for LW photon absorption  & 2.2\\
      $N_F$             & Gas column density for LW  absorption & 2.2\\
      $F_{\rm [CII]}$   & [CII] line flux emerging from the cloud & 3\\
      \Scii             & [CII] surface brightness & 5\\
      \S*               & Star formation rate per unit area & 5\\
      $\Sigma_g$        & Gas surface density & 5\\
      $\sigma$          & Gas r.m.s. turbulent velocity  & 5\\
      $k_s$             & Burstiness parameter  & 5\\
      \end{tabular}
\end{table}
\section{Introduction}
Constraining the properties of the interstellar medium (ISM) in the first galaxies that formed during the Epoch of Reionization (EoR) is a fundamental step to understand galaxy evolution and its impact on the reionization process \citep[for a recent review see][]{Dayal18}. In the last five years, the advent of ALMA started revolutionising the field of ISM studies at high-$z$. ALMA has allowed to detect with high resolution and sensitivity line emission tracing the cold gas phases (neutral and molecular) of the ISM in normal star-forming galaxies (${\rm SFR}<100\, \rm {M_{\odot} {\rm yr}^{-1}}$) at $z>6$, that are representative of the bulk of galaxy population at the end of EoR  \citep[e.g.][]{maiolino2015, capak2015,willott2015,knudsen2016, inoue2016, pentericci2016, matthee2017, bradac2017, carniani2017, jones2017, carniani2018, carniani2018b, smit2018, moriwaki2018, Tamura2018, Hashimoto2018}. 

Among all possible line emissions falling in the ALMA bands from $z>6$, the $^{2}P_{3/2}\rightarrow^{2}P_{1/2}$ forbidden transition of singly ionised carbon ([\ion{C}{ii}]) at $158\mum$ represents the workhorse for ISM studies. It is typically the most luminous line \citep{stacey1991} in the far-infrared (FIR), and it provides unique information on neutral and ionised phases associated with dense photo-dissociation regions (PDR) in the outer layers of molecular clouds \citep[][]{hollenbach1999, wolfire2003}.

A tight relation between \CII line luminosity and global star formation rate (SFR) is found from local galaxy observations \citep{delooze2014, Herrera2015}. However, the behavior of \CII line emission at $z>5$ appears much more complex than observed locally. ALMA observations have shown that only a sub-sample of the available \CII detections in early galaxies follows the  \citet{delooze2014} relation. The majority of high-$z$ sources, though, presents a large scatter around the local relation, with a considerable (15-20\%) fraction of the outliers being \quotes{[\ion{C}{ii}]-deficient} with respect to their SFR \citep[e.g.][]{carniani2018}.  High-$z$ galaxies are characterised by large surface star formation rates ($> 1 M_\odot~{\rm yr}^{-1}~{\rm kpc}^{-2} $); their [CII] emission appears to be considerably more extended than the UV \citep{carniani2018, Fujimoto19}. As a result their [CII] surface brightness is much lower than expected from the relation derived from spatially resolved local galaxies. 

In the last years, both experimental \citep[e.g.][]{maiolino2015,capak2015, knudsen2016, matthee2017} and theoretical studies \citep{vallini2015, olsen2017, pallottini2017b, lagache2018, popping2016, popping2019} have concentrated on this issue. \citet{vallini2015} suggested that the fainter \CII line luminosity can be explained if these sources deviate from the Kennicutt-Schmidt relation and/or they have a low metallicity. These authors found that \CII emission substantially drops for metallicities $Z<0.2 \,Z_{\odot}$, a result later confirmed by \citet[][]{olsen2017,lagache2018}.

Beside low metallicities, high-$z$ galaxies also show evidence for peculiarly intense radiation fields \citep[e.g.][]{Stark2017}, and compact sizes \citep[][]{Shibuya2019}.  In this situation, radiative feedback (e.g. due to radiation from massive stars) can be more effective, causing for instance the photo-evaporation of molecular clouds \citep{gorti2002,decataldo2017,decataldo:2019}. This process indirectly regulates the line luminosity resulting from the associated photo-dissociation regions \citep{vallini2017}. Similar arguments have also been invoked to explain the \CII deficit in some local galaxies \citep{Herrera2018, Diaz2017}. Stronger/harder radiation fields may also alter the ionisation state of carbon atoms.  

Given the complex interplay of the various effects governing \CII emission from galaxies, it seems worth investigating in detail the physics of emission and its relation to the global galaxy properties. To this aim we have developed an analytical model which has been validated against numerical results. The model successfully catches the key emission physics, and can therefore be used in a straightforward manner to interpret available and future \CII data both from local and high redshift galaxies. The study is similar in spirit to e.g. \citet{Munoz2016,Narayanan2017}, albeit these models concentrate on more specific aspects of the problem. 

\begin{figure*}
\includegraphics[scale=0.78]{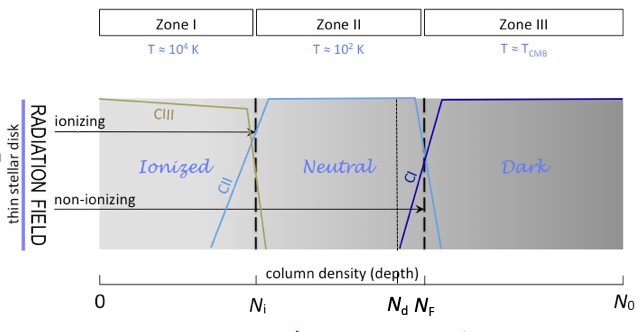}
\caption{Schematic ionisation structure of slab in our galaxy sandwich model. The radiation field produced by the thin layer of stars located at the galaxy mid-plane illuminates the overlying gas slab of total column density $N_0$. Zone I, extending up to $N_i$, is ionised and C is mostly in \CIIIion form. In Zone II only non-ionising ($h\nu < 1 Ryd$) FUV photons penetrate; nevertheless these can keep carbon in \CIIion form. We also highlight the column density $N_d$ at which the dust optical depth to non-ionising UV photons becomes equal to unity. Finally, beyond $N_F$ even FUV radiation is totally absorbed and carbon is neutral.     
\label{fig:sketch_gas_slab} }
\end{figure*}

\section{Ionisation structure}\label{sec:method}
Suppose that a galaxy can be modelled as a plane parallel slab of gas with uniform number density $n$, metallicity $Z$, total gas column density $N_0$, illuminated by a source emitting both ionising (photon energy $h_P\nu > h_P \nu_L = 13.6$ eV) and non-ionising radiation with specific energy flux ($\rm erg\, \rm cm^{-2} \rm s^{-1} \rm Hz^{-1}$)
\be
F_\nu = {F_L} \left(\frac{\nu}{\nu_L}\right)^{-\beta},  
\label{F}
\ee
where $F_L$ is the flux at the Lyman limit $\nu_L=3.2\times 10^{15}$ Hz, $\beta=4-5$ for normal Pop II stars, and $\beta \simeq 1.5$ for AGN-like sources. For now, we consider $N_0$ as a parameter of the model, in Sec. \ref{todata} we will connect it with other galaxy properties. 

Fig. \ref{fig:sketch_gas_slab} portraits a sketch of the ionisation structure of the slab. The ionising photons create a \HII region (Zone I) in which carbon is mostly in \CIIIion form, extending up to a column density $N_i$. Beyond $N_i$ the gas becomes neutral (Zone II), but non-ionising UV photons maintain carbon in a singly ionised state. The neutral layer extends up the point at which UV photons are absorbed by dust and H$_2$ molecules, at a column density $N_F$ which slightly exceeds $N_d$ (see derivation in Sec. \ref{sec:NF_calculation}), where the optical depth due to dust reaches unity. At even larger depths (Zone III) the gas is UV dark, and the only heating is provided by cosmic rays (and CMB at high redshift). In this region the gas is mostly in molecular form and carbon is found in a neutral state. In the following we characterise in detail Zone I and Zone II from which \CII line emission is produced.

\subsection{Ionised layer (Zone I)}
 The integrated ionising photon flux (cm$^{-2}\rm s^{-1}$) impinging on the cloud is then
\be
F_{i} = \int_{\nu_L}^\infty   \frac{F_\nu}{h_P \nu} d\nu = \frac{F_L}{h_P \beta}.
\label{def_ionizing_flux}
\ee
It is convenient to introduce the ionisation parameter, $U$, defined as the ionising photon-to-gas density ratio: 
\be
U = \frac{n_\gamma}{ n} =  \frac{F_{i}}{nc},
\label{U}
\ee
which implies $F_{i}=Unc$. We also define the Str\"omgren depth, 
\be
l_S =  \frac{F_{i}}{n^2 \alpha_B} =  \frac{U c} {n \alpha_B}, 
\label{xS}
\ee
where $\alpha_B =2.6\times 10^{-13} {\rm cm}^3 \rm s^{-1}$ is the Case-B recombination coefficient at temperature $T \approx 10^4$ K and $n$ is the gas density. The corresponding column density is 
\be
N_S = n l_S =  \frac{U c}{ \alpha_B} \approx 10^{23} U\,\, \rm cm^{-2}. 
\label{NS}
\ee
Then, we can write the photo-ionisation rate as
\be
\Gamma= \int_{\nu_L}^\infty   \frac{F_\nu e^{-\tau}}{ h_P \nu} \sigma_\nu d\nu = \frac{F_L\sigma_L}{h_P(3+\beta)}  e^{-\tau} \equiv \bar \sigma F_{i} e^{-\tau},
\label{Gamma}
\ee
where the hydrogen photo-ionisation cross-section is $\sigma = \sigma_L (\nu/\nu_L)^{-3}$, with $\sigma_L = 6.3\times 10^{-18} \rm cm^2$, and we have defined the flux-weighted photo-ionisation cross section as $\bar\sigma = \beta\sigma_L /(3+\beta)$ 

Then we apply the condition for photo-ionisation equilibrium to derive the hydrogen ionisation fraction, $\fhii=n_e/n$ (for simplicity we neglect He) as a function of depth, $l$, in the slab. This reads
\be
\frac{1-\fhii}{ \fhii^2} = \frac{\alpha_B n}{ \Gamma} = \frac{e^\tau}{\tau_s},
\label{nHI}
\ee
where $\tau_s= n \bar\sigma l_S$, and we have used eqs. \ref{xS}-\ref{Gamma} to obtain the last equality. The total optical depth $\tau$ appearing in eq. \ref{Gamma} includes both photoelectric and dust absorption. The flux-weighted dust extinction cross section per H-atom is $\bar\sigma_d =  5.9\times 10^{-22}  {\cal D}\, \rm cm^2$, where the dust-to-gas ratio ${\cal D}$ (in units of the Milky Way value) is taken to be equal to the gas metallicity in solar units.
Note that this simple assumption might break down at low metallicities as discussed by \citet{Remy14} and \citet{deCia13}.

Given the uncertainties in the optical properties of dust in \HII regions, it is reasonable to ignore the difference in the cross sections for ionising and non-ionising photons. We will also ignore the effects of radiation pressure on dust in the determination of the properties of the ionised layer \citep{Draine11}. 

The radiative transfer equation yields $\tau$ as a function of depth in the slab:
\be
\frac{d\tau}{ dl} = (1-\fhii) n \bar\sigma + n \bar\sigma_d.
\label{RT1}
\ee
From now on we will assume that the source of radiation are Pop II stars, and therefore set $\beta=4$. Hence,
\begin{subequations}\label{2tau}
\begin{align}
&\tau_s = n \bar\sigma l_S = 2.7\times 10^5 U, \\
&\td   =n \bar\sigma_d l_S \equiv N_S/N_d = 59\, U {\cal D},
\end{align}
where
\be\label{def_n_d}
N_d= 1/\bar \sigma_d = 1.7\times 10^{21}  {\cal D}^{-1} \,\rm cm^{-2}
\ee
is the gas column density at which the dust optical depth to UV photons becomes equal to unity. 
\end{subequations}
\begin{figure}
\includegraphics[scale=0.37]{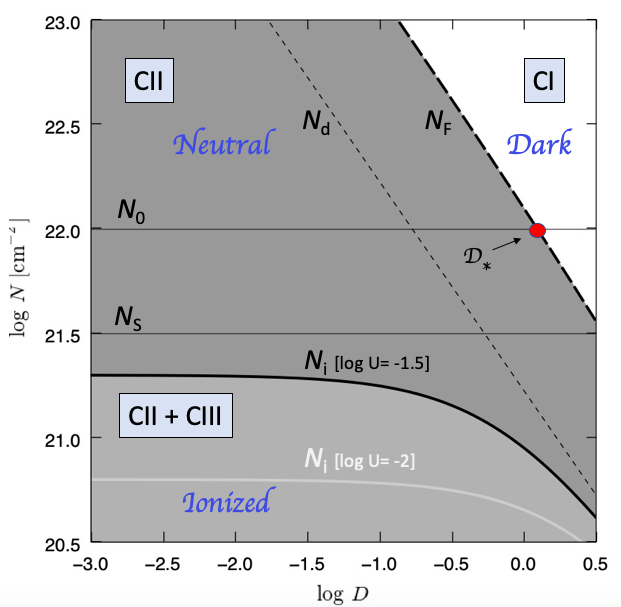}
\caption{Mutual relations among the key column densities ($N_S$, $N_F$, $N_i$,and $N_0$) as a function of the dust-to-gas ratio ${\cal D}$ and $\log U=-1.5$, showing the presence of ionised regions (containing both CII and CIII), neutral regions (CII only) and dark regions shielded from UV light in which neutral carbon (CI) is found. The figure assumes that the slab has a total column density $\log N_0=22$. For comparison, the $N_i$ curve for $\log U=-2.0$ is also shown. We also highlight the dust to gas ratio, $D_*$, for which $N_F=N_0$.
As a reference, $N_S$, $N_D$, $N_i$, and $N_F$ are given in eq.s \ref{NS}, \ref{def_n_d}, \ref{def_NI}, and \ref{NF}, respectively.
}
\label{fig_regimes}
\end{figure}
%
It is convenient to introduce the dimensionless depth $y=l/l_S$. Eq. \ref{RT1} then becomes
\be
\frac{d\tau}{ dy} = (1-\fhii) \tau_s + \td.
\label{radtrans}
\ee
By substituting eq. \ref{nHI} into eq. \ref{radtrans}, and taking $1-\fhii \ll 1$ (a condition well satisfied in the ionised region) we get
\be
\tau(y) = -\ln\left[\left(1+ \td\right) \frac{e^{-\td y}}{ \td } - \frac{1}{ \td}\right] ,
\label{tau}
\ee
which satisfies the boundary condition $\tau(y=0)=0$. By using eq. \ref{nHI}, the final solution for the ionised fraction\footnote{For $\td \ll 1$ eq. \ref{ionfrac} reduces to ${1/(1-\fhii)} = \tau_s (1 - y -\td y)$. As this approximation requires $N_S/N_d \ll 1$, it is valid only for $U\ll 0.017 {\cal D}^{-1} $, i.e. relatively small \HII layers.} in the slab is
\be
\frac{1}{1-\fhii} = 
\frac{1}{\fhi} =\left(\frac{\tau_s}{\td}\right) \left[\left(1 + \td\right) {e^{-\td y}} - 1\right]. 
\label{ionfrac}
\ee

In analogy with $N_d$ (eq. \ref{def_n_d}), we define the depth of the \HII region, $y_i$ such that $\tau(y_i) =1$. 
This yields 
\be
y_i = {1\over \td} \log \left({1+\td \over 1+\td/e}\right). 
\label{yi}
\ee
Recalling that $y_i =N_i/N_S$, and $\td = N_S/N_d$, it follows that 
\be
N_i = y_i N_S = N_d \ln \left(\frac{1+\td}{ 1+\td/e}\right). 
\label{def_NI}
\ee
When dust opacity is small ($\td \rightarrow 0$) we get $y_i =(1-1/e)$, or equivalently $N_i = 0.63 N_S$; hence, the previous expression almost correctly converges to the classical, dust-free Str\"omgren solution. The small discrepancy is due to the assumption  $1-x \ll 1$ made to solve eq. \ref{tau}.
In the opposite, optically thick regime (e.g. high ${\cal D}$ or $U$ values) $N_i \simeq N_d$. This implies that \HII regions cannot extend beyond $N_d$. 

The mutual relations among the various quantities so far introduced are shown in Fig. \ref{fig_regimes}; we will physically interpret them shortly. 
For now we notice that depending on the relative values of  $N_i$ and $N_0$, the \HII region can be either ionisation- ($N_0 > N_i$) or density-bounded ($N_0 < N_i$).

It is useful to derive also the total \HI column density in the ionised layer by integrating eq. \ref{ionfrac} from $y=0$ to $y_i$ and using the expression for $y_i$ given by eq. \ref{def_NI}:
\begin{align}
N_{\rm HI}(y_i) = \, & N_S \int_0^{y_i} \fhi dy = \frac{N_S}{\tau_s} \ln \frac{\td}{ \vert e^{\td y_i}-\td -1\vert} \nonumber\\ 
=\, & \frac{N_S}{ \tau_s} \ln \frac{e(1+\td/e)}{ 1+\td} =  \frac{N_S} {\tau_s} \left( 1- \frac{N_i}{N_d}\right).
\label{NHI}
\end{align}
The last equality implies that the neutral to ionised column density ratio within the \HII region is very small,
\be
\frac{N_{\rm HI}}{N_i} = \frac{\td}{\tau_s}\left(\frac{N_d}{N_i} -1 \right) \approx \frac{\cal D}{4576} \left(\frac{N_d}{N_i} -1 \right) \ll 1,
\ee
and it is decreasing for larger $U$ values.

Inside Zone I carbon is mostly in the form of \CIIIion. In principle, one can compute the relative abundance of doubly and singly ionised species (see Appendix \ref{appendix:c_ionization}). However, we show below that in ionisation-bounded regimes the contribution to \CII emission from the \HII region is almost completely negligible in all cases of interest. Hence to keep things simple, we assume that $n_{\rm CII}/n_{\rm CIII} \approx \fhi$.

\subsection{Neutral layer (Zone II)}
\label{sec:NF_calculation}
The extent of the \CIIion layer (Zone II in Fig. \ref{fig:sketch_gas_slab}) is set by the penetration of non-ionising UV photons in the energy range $11.26-13.6$ eV, corresponding to $\lambda=912-1102$ \AA, which are capable of singly ionising carbon atoms ($I_C=11.26$ eV) beyond $N_i$. The above energy range coincides almost precisely with the Lyman-Werner (LW) H$_2$ dissociation band ($912-1108$ \AA), so in what follows we neglect such tiny difference; for simplicity we refer to both bands as the LW band. 

The frequency-integrated photon flux ($\rm cm^{-2}\, s^{-1}$) in the LW band at the slab surface is given by (see eq. \ref{F}):
\begin{equation}
F_0 = \int_{\nu_1}^{\nu_L} \frac{F_{L}}{h_P\nu} \left(\frac{\nu}{\nu_L}\right)^{-\beta} d\nu\approx 0.3 h_P^{-1}F_L
\end{equation}
where $h \nu_1=11.26$ eV. Recalling eq. \ref{def_ionizing_flux}
it follows that the impinging FUV flux is
\begin{equation}
F_0= 0.3 \beta F_i = 0.3 \beta U n c = 3.6\times 10^{10} U n
\label{F0}
\end{equation}

The decrease of LW photon flux inside the slab is regulated by the radiative transfer equation \citep[see e.g.][]{sternberg2014}:
\begin{equation}
\frac{dF}{dl}= -n \bar\sigma_{d} F - \frac{{\mathcal R} n \nhi}\fd 
\label{rt_LW}
\end{equation}
where the first (second) term on the r.h.s. of eq. \ref{rt_LW} accounts for dust (H$_2$ line) absorption of LW photons. In Zone II, the total density $n=\nhi+2 \nhh$ is the sum of the atomic ($\nhi$) and molecular ($\nhh$) hydrogen density;  $\fd \approx 0.15$ \citep{krumholz2008} is the fraction of absorbed LW photons leading to \HH dissociations, and ${\mathcal R}\approx 3\times 10^{-17} T_2^{1/2} {\mathcal D}\, \rm cm^{3} s^{-1}$, is the rate\footnote{We use the compact notation $Y_x = Y/10^x$} of H$_2$ formation on dust grains \citep{cazaux2004}. Clearly, if \HH formation is not considered (${\mathcal R}=0$), eq. \ref{rt_LW} yields the standard dust attenuation solution $F = F_0 \exp{(-N/N_d)}$. Rewrite Eq. \ref{rt_LW} as:
\begin{equation}
\frac{d\mathcal F}{d\tau} = - \mathcal F - \frac{\fhi}{\chi},
\label{rt_LW1}
\end{equation}
where ${\mathcal F} \equiv {F}/{F_0}$, $\tau = n \bar\sigma_{d} l$, and we have introduced the parameter
\begin{equation}
\chi = \frac{\fd \bar\sigma_d F_0}{{\mathcal R}n}.
\label{chi_parameter}
\end{equation}
Then, also using the expression for $F_0$ in   eq. \ref{F0}, 
we obtain 
\begin{equation}
\chi = 2.9 \times 10^{-6} \frac{F_0}{n} = 10^5\, U.
\end{equation}

In reality, it is necessary to account for the probability that a LW photon is absorbed by dust grains (associated with H$_2$) rather than by H$_2$ molecules in Zone II. This effect is embedded in the extra factor $w$ \citep{sternberg2014} defined as
\begin{equation}
    \begin{split}
    w & \equiv \frac{W_{g,tot}}{W_{d, tot}} = \frac{\rm effective \,bandwidth\, for\, dust \,abs.}{\rm effective\, bandwidth\, for\, H_2 \,abs.} =\\ &= \frac{1}{1 + (\bar\sigma_d/7.2 \times 10^{-22} \rm cm^2)^{1/2}};
    \end{split}
\end{equation}
for the presently assumed value $\bar\sigma_d = 5.88 \times 10^{-22} \mathcal{D}$, we obtain
\begin{equation}\label{eq_w_dust}
w = \frac{1}{1+0.9 \mathcal{D}^{1/2}}. 
\end{equation}
Hence, if we replace $\chi$ with $\chi'= w \chi$ in eq. \ref{rt_LW1}, and use the above definitions, we obtain
\begin{equation}
\frac{d(wF)}{dl} = -\nhi \bar\sigma_d w F - \frac{R n \nhi}{\fd}.
\label{wF_eq}
\end{equation}
Note that in the dust absorption term we replaced $n$ with $\nhi$ because the absorption due to dust associated with \HH has been now included into $w$; this entails also the redefinition of  $\tau_{\rm HI} = \nhi \bar\sigma_{d} l$.
Eq. \ref{wF_eq} can be simplified as
\begin{equation}
 \frac{d\mathcal F}{d\tau_{\rm HI}} = - \mathcal F - \frac{\fhi}{\chi'}.
 \label{F_final_eq}
\end{equation}
If we further assume that the atomic fraction $\fhi= 1$  everywhere $F$ is non-zero (this is equivalent to assuming a sharp transition), we get the final solution,
\begin{equation}
{\mathcal F}(\tau) = \frac{\chi' +1}{\chi'} e^{-\tau} - \frac{1}{\chi'}.
\end{equation}

From now on we indicate with $N_F$ (see Fig. \ref{fig:sketch_gas_slab}) the column density at which the LW flux vanishes, i.e. ${\mathcal F}(\tau) = 0$. This happens at $\tau_F = {\ln}(1+\chi')$, with $\tau_F=N_F/N_d$. Hence the column density $N_F$ can be written:
\begin{equation}
N_F = N_d \ln(1+ \chi ') = N_d \ln(1+ 10^5 w U).
\label{NF}
\end{equation}

Fig. \ref{fig_regimes} helps elucidating the mutual relations among the fundamental scales of the problem as a function of ${\cal D}$ at fixed $\log U = -1.5$, although for comparison we show also $N_i$ for $\log U=-2.0$. For low dust-to-gas ratios, $N_F$ largely exceeds the \HII region column density $N_i \approx 10^{21.3} \rm \, cm^{-2}$. Inside the \HII region (referred to as Zone I in Fig. \ref{fig:sketch_gas_slab}) carbon is largely in the form of \CIIIion; beyond $N_F$ carbon becomes neutral (Zone III). When the dust-to-gas ratio reaches  ${\cal D} = {\cal D}_*$, $N_F$ drops below $N_0$, here assumed to be equal to $10^{22} \rm cm^{-2}$ for display purposes. As shown later, the position of ${\cal D}_*$ marks a distinctive change in the \CIIion emission.  Finally, as the dust abundance increases beyond ${\cal D}_*$ both the size of the \HII region and the thickness of the neutral layer shrink: the layer becomes \CIion--dominated. 

\section{Emission model}\label{sec_emission_model}
The \CII line flux emitted by a slab with total gas column density $N_0$ depends on whether the \HII region is ionisation- or density-bounded. We assume that the \HII layer (Zone I) has a temperature $T=10^4$ K, whereas in Zone II we set $T=10^2$ K. We will return on the impact of this assumption when validating the model in Sec. \ref{Val}.
The abundance of carbon is taken to be ${\cal A_C}=2.7\times 10^{-4}$  \citep{asplund2009}, and we linearly scale it with ${\cal D}$, or equivalently, given our assumption of a constant dust-to-metal ratio, with $Z$. 
In general, the \CII line flux (erg  cm$^{-2} \rm s^{-1}$) emerging from the slab is given by
\be
F_{\rm [CII]} =  n_x n_j \Lambda_j(T) \ell_0 = n_x N_j \Lambda_j(T) \label{flux}
\ee
where $n_x$ ($n_j$) is the number density of collisional partner (target ion) species, $\Lambda_j(T)$ are the appropriate cooling functions (see the derivation in Appendix \ref{appendix:cooling_functions}), and $\ell_0=N_0/n$.

\subsection{Ionisation-bounded regime}\label{sec_ion_bound}
\begin{figure*}
    \centering
    \includegraphics[scale=0.24]{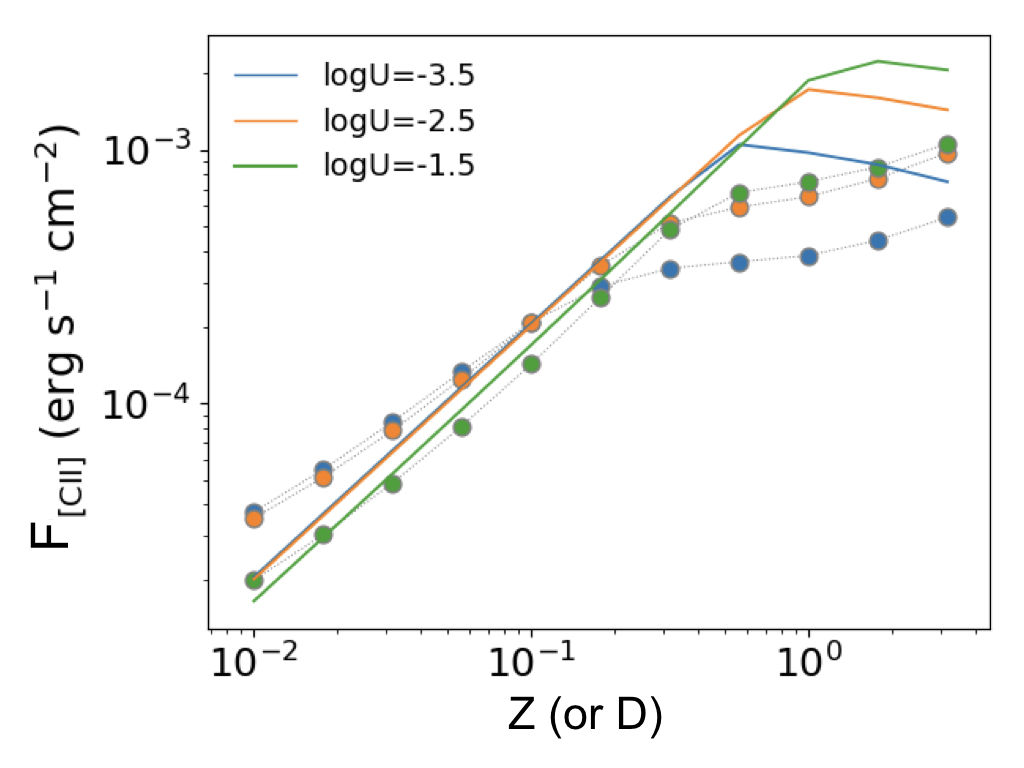}
     \includegraphics[scale=0.24]{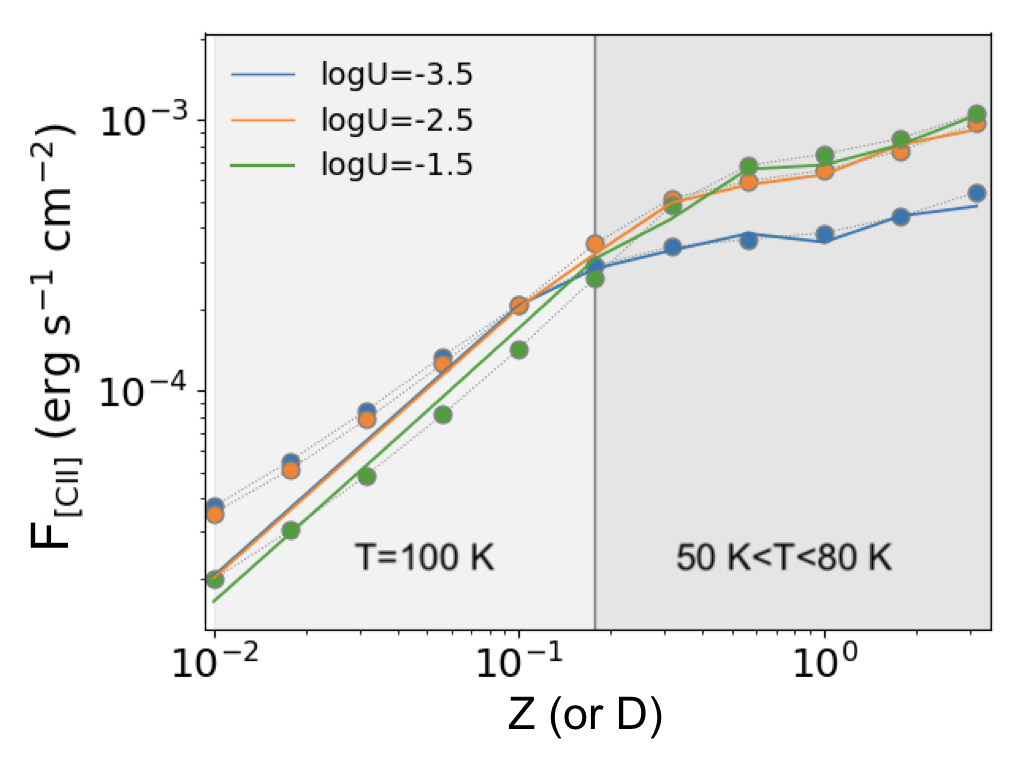}
    \caption{{\it Left panel}: [CII] line flux emerging from  slab with $N_0=10^{22} {\rm cm}^{-2}, n=100\, \rm cm^{-3}$ as a function of $Z$ (or equivalently, $D$, both in solar units) for three different values of the ionisation parameter: $\log U=-1.5$ (green line), $\log U=-2.5$ (orange), and $\log U=-3.5$ (blue). We assume a constant neutral layer temperature  $T=100 \, \rm K$. The points (colour coded according the same convention) represent the results obtained from \code{cloudy} simulations. {\it Right}: Same as the left panel, however different gas temperatures are assumed in the two metallicity ranges $Z \gtrless 0.18$ for different ionization parameters.  The model assumes $\langle T\rangle=58 \rm \, K$ for $\log U=-3.5$, $\langle T\rangle=68 \rm \, K$ for $\log U=-2.5$, and $\langle T\rangle=75 \rm \, K$ for $\log U=-1.5$.}
    
    \label{fig:CII_emissivity}
\end{figure*}
Let us consider first the most likely case of ionisation-bounded \HII regions ($N_i < N_0$).
For hydrogen column densities $N_H < N_i$ the carbon is essentially all in the form of \CIIIion, and therefore $n_{\rm CIII} \approx n_C = {\cal A_C D}n$ (we neglect the possible presence of \CIVion). However, traces of \CIIion are present also in the ionised region. Hence, in principle, one should compute the exact C ionisation fraction. In Appendix \ref{appendix:c_ionization} we show that the singly ionised carbon fraction $x_C$ closely follows the hydrogen ionisation fraction $\fhi$ derived in eq. \ref{ionfrac}. Given that the \CII emission contributed by Zone I is negligible in the ionisation-bounded regime, for simplicity we adopt the approximation $n_{\rm CII} = \fhi n_{\rm C}$.

Beyond $N_i$ the gas becomes neutral and carbon is maintained in the singly ionised state by the LW radiation field penetrating the slab beyond the \HII region; hence $n_{\rm CII} \simeq n_C$. Then it turns out that in almost all cases the neutral region (Zone II) provides the dominant contribution to the total emitted \CII~ flux. This regions extends from $N_i$ (eq. \ref{def_NI}) to $N_F$ (eq. \ref{NF}), as shown in the previous Section.

The emerging [CII] line flux contains the contribution from the ionised and neutral layers. In the ionised layer the collision partner for \CIIion ions are electrons, $n_x\equiv n_e \approx n$, and $N_{\rm CII} \approx {\cal A_C D} N_{\rm HI}(y_i)$, where $N_{\rm HI}$ is obtained in eq. \ref{NHI}. Such emission must be augmented with the one arising from the neutral layer, where the collisional partners are H atoms, $n_x\equiv n_H \approx n$. As the extent of the neutral layer is limited by absorption of LW photons (see Section \ref{sec:NF_calculation}), we set $N_{\rm CII} = \min(N_F, N_0)-N_i$. It follows that 
\begin{equation}
\begin{split}
F_{\rm [CII]} &= n {\cal A}_{C} {\cal D} \left\{ \Lambda^{(4)}_{\rm [CII]} N_{\rm HI}(y_i) + \Lambda^{(2)}_{\rm [CII]} \left[ {\min}(N_F, N_0) - N_i \right] \right\},
\end{split}
\label{Fcii_ionbound}
\end{equation}
where $\Lambda^{(n)}=\Lambda(T=10^n \rm K)$. It is easy to show\footnote{The condition for the first term to become dominant is $N_{\rm HI} > (\Lambda^{(2)}_{\rm [CII]}/\Lambda^{(4)}_{\rm [CII]}) \min(N_0, N_F) = 0.006 N_0$. As from eq. \ref{NHI}, $N_{\rm HI} < 3.7 \times 10^{17}\, \rm cm^{-2}$, for $N_0$ (or $N_F$) $> 6 \times 10^{19}\, \rm cm^{-2}$ \CII emission from the ionised layer can be safely neglected to a first-order approximation.} that the first term, accounting for the ionised layer emission, is almost always sub-dominant in the ionisation-bounded regime.   

As an example, in Fig. \ref{fig:CII_emissivity} (left panel) we plot  $F_{\rm [CII]}$ from eq. \ref{Fcii_ionbound} for a slab with $N_0=10^{22}\, {\rm cm}^{-2}$ and $n=100\, \rm cm^{-3}$. Two regimes can be clearly identified. The first occurs when $N_0<N_F$, i.e. for dust-to-gas ratios ${\cal D}<{\cal D}_*$ (see Fig. \ref{fig_regimes}), where the flux can be approximated as
\be
F_{\rm [CII]} \approx n {\cal A}_{C} {\cal D} \Lambda^{(2)}_{\rm [CII]} \left[N_0 - N_i \right] \propto {\cal D} N_0\,;
\label{lowZapprox}
\ee
note that we neglected the \HII layer contribution and we have further assumed that $N_i \ll N_0$ in the last passage. In this regime the [CII] luminosity grows linearly with metallicity. In the high-metallicity regime (${\cal D}>{\cal D}_*$), the depth at which LW photons are fully absorbed becomes smaller than the thickness of the layer ($N_F < N_0$). It follows that (again neglecting the \HII layer contribution) \be
\begin{split}
F_{\rm [CII]} &\approx n {\cal A}_{C} {\cal D} \Lambda^{(2)}_{\rm [CII]} \left[N_F - N_i \right]\\ &\propto n {\cal D} N_d {\ln}(1+ 10^5 w U).
\label{hiZapprox}
\end{split}
\ee
In this regime the \CII flux has a weak, logarithmic, dependence $U$, and a weaker one on ${\cal D}$, as the product ${\cal D} N_d$ is constant (eq. \ref{def_n_d}), and $w \sim {\cal D}^{-1/2}$ (eq. \ref{eq_w_dust}).
This explains (i) the plateau seen in all curves in Fig. \ref{fig:CII_emissivity} for ${\cal D}>{\cal D_*}$, and (ii) its increasing amplitude with $U$. 
The transition between the two regimes is located at ${\cal D}={\cal D_*} \approx 0.3-1$ and seen as a kink in the curves. The kink shifts towards higher metallicities for larger values of $U$ as a result of the fact that, approximately, $D_* \propto \ln U$.

\subsection{Density-bounded regime}\label{sec_den_bound}

The density-bounded regime occurs if $N_0 < N_i$, when the slab becomes fully ionised. In this case
the \CII emitting column density can be computed in the same way as for the ionisation-bound case from the integral of the \HI density profile within the \HII region up to $y_0=\ell_0/l_s$. 
\be
N_{\rm HI}(y_0) =  N_S  \int_0^{y_0} \fhi dy = \frac{N_S}{\tau_s} \ln \frac{\td}{ \vert e^{\td y_0}-\td -1\vert}. 
\label{NHI_aty0}
\ee
The corresponding line flux follows from eq. \ref{Fcii_ionbound}, 
\be
F_{\rm [CII]} =  n {\cal A_C D} \Lambda_{\rm [CII]}^{(4)} N_{\rm HI}(y_0). 
\label{FC2D}
\ee

\section{Model validation}\label{Val}

We validate the predictions of the analytical model against accurate numerical simulations of Photo-Dissociation Regions. To this aim, we run a set of simulations using \code{cloudy} v17.0 \citep{ferland2017}. We consider a 1D gas slab with constant gas density, $\log(n/{\rm cm^{-3}})=2$ and total column density $N_0=10^{22}\rm \, cm^{-2}$. The gas metallicity\footnote{We assume solar abundances (\texttt{abundances GASS}) from \citet{grevesse2010} for which ${\cal A}_{C}=2.7 \times 10^{-4}$ as also assumed in the analytical model.} can vary in the range $\log Z=[-2.0, 0.5]$. As for the dust, we set the dust-to-metal ratio $\fdust = 0.3$, and assume ISM grains (\texttt{grains ISM}) with a size distribution, abundance and materials (graphite and silicates) appropriate for the ISM of the Milky Way \citep{mathis1977}.

The gas slab is illuminated by stellar sources with a spectral energy distribution (SED) obtained from the version v2.0 of the Binary Population and Spectral Synthesis (BPASS) models \citep{stanway2016}. Among the sample of BPASS models we select those with  $Z_{\star}= 0.5$. A broken power-law is used for the initial mass function (IMF), with a slope of $-1.3$ for stellar masses $m_{\star}/\msun \in [0.1,0.5]$ and $-2.35$ for $m_{\star}/\msun \in (0.5,100]$. We adopt a continuous star formation mode and select models at $10 \rm \, Myr$. We scale the SED to obtain ionisation parameters at the gas slab surface $\log U \in [-3.5, -2.5, -1.5]$. In Fig. \ref{fig:CII_emissivity} \code{cloudy} results are shown as coloured points.

\begin{figure}
    \centering
    \includegraphics[scale=0.5]{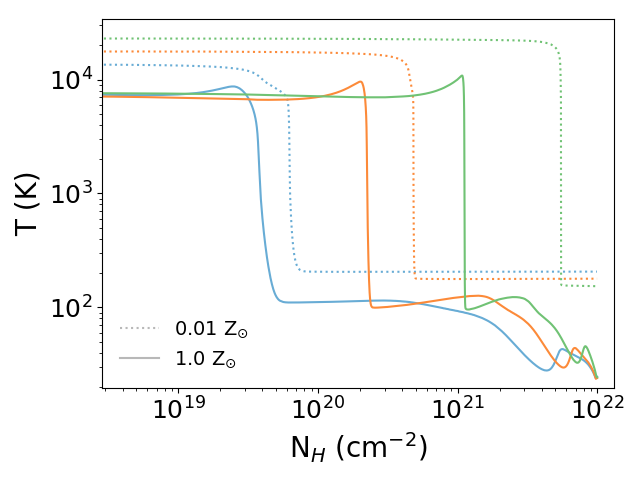}
    \caption{\code{cloudy} temperature profiles for a slab with $N_0=10^{22} {\rm cm}^{-2}, n=100\, \rm cm^{-3}$, and for two extreme metallicities: $Z=0.01 Z_{\odot}$ (dotted lines) and $Z=Z_{\odot}$ (solid lines). The colour code is the same as in Fig. \ref{fig:CII_emissivity}, $\log U=-1.5$ green, $\log U=-2.5$ orange, $\log U=-3.5$ blue lines. }
    \label{fig:temp_profiles}
\end{figure}

Given its simplicity, the analytical model is in striking overall agreement with \code{cloudy} results. In particular, the model correctly predicts both the amplitude and linear slope of the increasing flux trend at low metallicity. \code{cloudy} results also confirm the presence of a plateau at higher metallicities, along with the correct positive correlation of its level with increasing ionisation parameter values. Finally, the predicted rightwards shift of the kink for higher $U$ values is substantiated by the numerical results.

In spite of such general compliance, some discrepancies remain affecting the plateau region. These concern the (a) amplitude and slope of the plateau, and (b) the location of the kink point. Although the differences are relatively small (for example, the plateau amplitude is overestimated by the model at most by a factor $\approx 2$ at any $Z$), it is worth investigating in depth their origin.

After performing additional analysis and tests, we concluded that the small differences are due to the assumed constant gas temperature in the neutral layer (Zone II). Fig. \ref{fig:temp_profiles} quantitatively highlights this fact.
The temperature profiles in the slab depend on both $Z$ and $U$.
In the ionised region (Zone I) the temperature is almost flat around a value around $10^4$ K, with essentially no dependence on $U$; and a weak, inverse dependence on $Z$ which decreases the \HII region temperature from $15\times 10^3$ to $8\times 10^3$ K for $Z = 0.01 \rightarrow 1$. The temperature profiles are also remarkably flat within this region. This perfectly justifies the temperature-independent value of $\Lambda^{(4)}_{\rm [CII]}$ adopted here.

In Zone II the temperature drops to values $T\approx 100$ K set by UV photo-heating, and consistent with the one we assumed to compute the cooling function ($\Lambda^{(2)}_{\rm [CII]}$). However, two effects make this approximation less precise: (a) at temperatures close to the resonance energy of the \CII transition, $E_{12}/k = 91.92$ K, the cooling function is very sensitive to small temperature variations; (b) the temperature profiles are not perfectly constant within Zone II, and they show variations of about $\pm 20 K$ from the mean of 100 K.

To verify that the origin of the discrepancies between the model and \code{cloudy} in the plateau region are indeed due to temperature variations within the neutral layer, we use the average temperature returned by \code{cloudy} for each metallicity when computing $\Lambda_{\rm [CII]}(T)$ in the model. The results of this test are shown in Fig. \ref{fig:CII_emissivity} (right panel). While at low metallicity (as seen also from Fig. \ref{fig:temp_profiles}) using a fixed $T=100$ K provides a very good approximation, at $Z>0.1$ the mean temperature is in the range $T=50-80\, \rm K$.
Once this correction is implemented in the model, its predictions almost perfectly match \code{cloudy} results over the entire range of metallicities and ionisation parameters.

In principle, one could improve the model by using an energy equation accounting for the temperature variations in the neutral layer. However, this would make the model more complicated, and likely not treatable in analytical terms. In the following we will stick to this simplification. Of course, if very accurate estimates are required by a given problem, one can always resort to full \code{cloudy} simulations.

\section{Bridging model and data}\label{todata}
We now aim at interpreting the available data on \CII emission from galaxies, both at low and high redshift, using the model\footnote{Note that at the gas densities ($\simgt 100\cc$), and temperatures ($T> 100$ K) considered here, line suppression effects by the CMB are negligible \citep{pallottini2017}.} developed in the previous Sections. Our model can be applied to a variety of experiments and problems involving \CII measurements. As a first application, here we concentrate on the interpretation of the observed relation between \CII and star formation rate. As pointed out in the Introduction, this issue is at the core of ISM and galaxy evolution studies. Even more excitingly, newly acquired ALMA data make it possible to produce a decisive step in the study of the internal structure and the ISM of galaxies in the EoR.

In order to effectively compare the results of the model to the observed \CII line emission a few more steps are required\footnote{We compare our model with luminosity and SFR surface density measurements rather than with the corresponding galaxy-integrated quantities $L_{\rm CII}$ and SFR. This choice overcomes the uncertainties related with the determination of the \CII emitting and star forming radii, which might in addition differ \citep{carniani2018}. Of course, if information about the spatial extent is available all the quantities given here can be then straightforwardly transformed.}. First we need to convert the main output of the model, the emitted flux  $F_{\rm [CII]}$ (${\rm erg}\,{\rm cm}^{-2}\,{\rm s}^{-1}$), in the more standard units of surface luminosity, \Scii ($L_\odot$ kpc$^{-2}$). The conversion factor is
\be
\Sigma_{\rm [CII]} = 2.4 \times 10^9  F_{\rm [CII]}\quad L_\odot$\, kpc$^{-2},
\label{eq_sigma_cii}
\ee
where $F_{\rm [CII]}$ is given by eq. \ref{Fcii_ionbound} or eq. \ref{FC2D} if the \HII layer is ionisation-bounded ($N_i < N_0$) or density-bounded ($N_i > N_0$), respectively.
The next step is to express $U$ (entering the flux expression via $N_F$ and $N_i$) and $N_0$ with respect to the observed quantity, the star formation rate per unit area $\Sigma_{\rm SFR}$, for which we adopt the standard units of $ M_\odot~{\rm yr}^{-1}~{\rm kpc}^{-2}$. 

Let us start with the average ionisation parameter, $U=\bar n_\gamma/\bar n$, where the bar indicates galaxy-averaged quantities. {Using the population synthesis code {\tt STARBURST99\footnote{\url stsci.edu/science/starburst99}}}, the ionising photon flux associated with $\Sigma_{\rm SFR}$ can be written as $\Sigma_\gamma = 3\times 10^{10} \Sigma_{\rm SFR} = \eta \Sigma_{\rm SFR}$ phot s$^{-1}$ cm$^{-2}$ for the stellar population properties assumed  in Sec. \ref{Val}, i.e. $Z_*=0.5$ and age 10 Myr. It follows that
\be
\bar n_\gamma = c^{-1} \Sigma_\gamma = \frac{\eta}{c}\Sigma_{SFR} \simeq \Sigma_{SFR}.
\label{ngamma}
\ee
The mean gas density is simply written in terms of the gas surface density, $\Sigma_g$ (in $M_\odot \,{\rm kpc}^{-2}$), and scale height, $H$, of the gas in the gravitational potential of the galaxy as follows:
\be 
\bar n=\frac{\Sigma_g}{\mu m_p H} = \frac{\pi G \Sigma_g^2}{\mu m_p \sigma^2} = 5.4 \times 10^{-13} {\Sigma_g}^2 \sigma_{\rm kms}^{-2} \quad \cc.
\label{mean_n}
\ee
In the previous equation, $\mu$ is the mean molecular weight of the gas (for simplicity we take it equal to 1), $m_p$ is the proton mass, and we have imposed the hydrostatic equilibrium to obtain the expression $H=\sigma^2/\pi G \Sigma_g$, where $\sigma$ is the gas r.m.s. turbulent velocity. For the latter quantity we assume the reference value $\sigma_{\rm kms} = \sigma/({\rm km\, s}^{-1}) = 10$ \citep{pallottini2017b,vallini2018}.

Note that in the following we will take $N_0 = \bar n H = \Sigma_g/\mu m_p$ using the definition in eq. \ref{mean_n}. This entails the implicit assumption that the emitting region (physically corresponding to a molecular cloud) column density, $n \ell_0$, is the same as the galaxy column density, $N_0$, implying $\ell_0 = \Sigma_g/\mu m_p n$, where $n=100\, \cc$ is the molecular cloud density adopted here. For Milky Way values ($\Sigma_g \approx 2\times 10^7 M_\odot\, {\rm kpc}^{-2}$), it is $\ell_0 \approx 10$ pc, which approximates well the typical observed size of Galactic molecular complexes.

We can now write a simple expression for the ionisation parameter,
\be
U = 1.7\times 10^{14}\,\frac{\Sigma_{\rm SFR}}{\Sigma_g^2}.
\label{Udef}
\ee
By noting that $\Sigma_g = \mu m_p N_0$, one can relate it to the slab column density as $\Sigma_g = 7.5\times 10^7 N_{0,22} M_\odot \rm kpc^{-2}$.

\begin{figure*}
    \centering
    \includegraphics[scale=0.65]{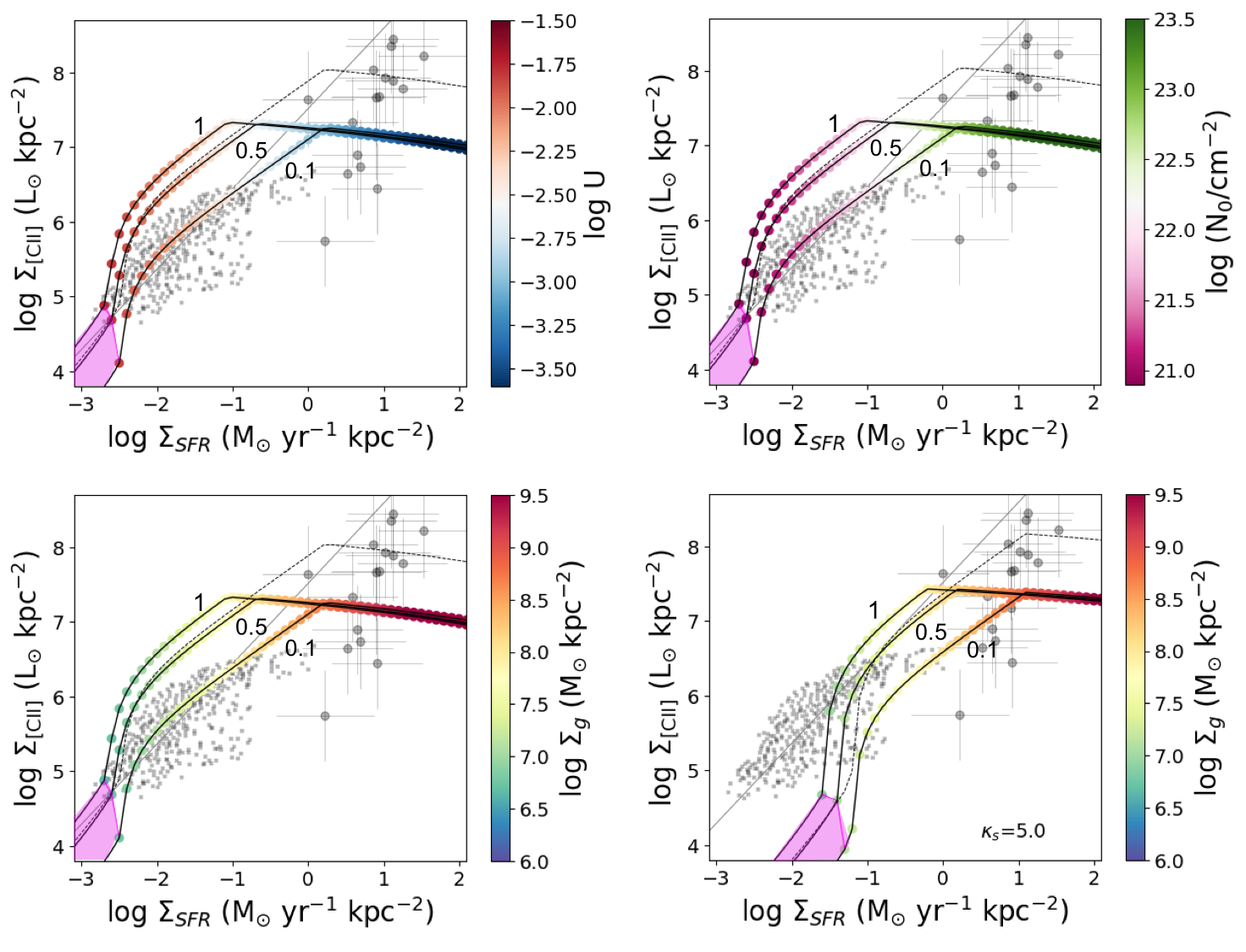}
    \caption{Predicted \Scii - $\Sigma_{\rm SFR}$  relation for three metallicities $Z=0.1, 0.5, 1$ (indicated by the labels), and gas density $n=500\, \rm cm^{-3}$. For reference, the thin dashed line shows the case $Z=0.1$ and $n=n_{\rm crit}=3027 \,\cc$. Model points are colour-coded according to values of ionisation parameter in panel (a), column density (b), and $\Sigma_g$ (c-d). In panel (d) we set $\ks=5$ (see eq. \ref{KS}); this value is typical of starburst galaxies \citep[e.g.][]{daddi2010}.  Gray small crosses are the \citet{delooze2014} data for a sample of spatially resolved local dwarf galaxies; the best fit to the data (eq. \ref{dLfit}) is shown by the thin solid line. Points with errors are either single galaxies without any sub-component, and data for individual components within galaxies taken from the sample of $z>5$ sources analized by \citet{carniani2018}. The magenta shaded area marks the density-bounded regime at different metallicities.}
    \label{fig:SigmaCIISigmaSFR}
\end{figure*}
To proceed further we need to eliminate $\Sigma_g$ from eq. \ref{Udef}. To this aim we use the empirical Kennicutt-Schmidt (KS) average relation between these two quantities, as given, e.g. by \citet{Heiderman10},
\be
\Sigma_{\rm SFR}= 10^{-12} \ks {\Sigma_g}^m \quad\quad (m=1.4),
\label{KS}
\ee
where we have allowed for deviations from the relation through the \quotes{burstiness} parameter $\ks$. 
Values of up to $\ks = 100$ have been measured for sub-millimeter galaxies (see e.g. \citet{Hodge15}.
Galaxies with $\ks > 1$ show a larger SFR per unit area with respect to those located on the KS relation having the same value of $\Sigma_g$ \citep{Hodge2015}. As we will see in the following Section, this parameter plays a crucial role in the interpretation of the \Scii - $\Sigma_{\rm SFR}$ relation. By inverting eq. \ref{KS} and eliminating $\Sigma_g$ from the expression for $U$, we finally get 
\be
U = 1.7\times 10^{(14 -24/m)} \,\ks^{2/m}\Sigma_{\rm SFR}^{(m-2)/m}  \simeq 10^{-3} \, \ks^{10/7} \Sigma_{\rm SFR}^{-3/7}.
\label{Udef1}
\ee
An interesting conclusion from the previous equation is that, perhaps contrary to naive expectations,  galaxies with larger $\Sigma_{\rm SFR}$ have a {\it lower} ionisation parameter, if they lie on the KS relation. However, starburst galaxies ($\ks > 1$) are characterised by higher $U$ values for the same $\Sigma_{\rm SFR}$. 

\section{Interpreting the \CII - SFR relation}\label{sec_cii_sfr_relation}

We now turn to the interpretation of the \Scii - $\Sigma_{\rm SFR}$ relation. We also compare our model to measurements of such relation in local ($z \approx 0$) and high-redshift ($z>5$) galaxies.
The local observations have been carried out by \citet{delooze2014} for a sample of spatially resolved low-metallicity dwarf galaxies using  the {\it Herschel} Dwarf Galaxy Survey. These authors provide the following fit:
\be
\log \Sigma_{\rm SFR} = -6.99 + 0.93 \log\Sigma_{\rm [CII]}.
\label{dLfit}
\ee
The relation is very tight, with an estimated 1$\sigma$ dispersion of 0.32 dex, and it implies that, at least locally, more actively star forming galaxies are brighter \CII emitters.

For the high redshift sample we use the recent determination by \citet{carniani2018}. These authors have used new ALMA observations of galaxies at $z =6-7$ as well as a re-analysis of archival ALMA data. In total 29 galaxies were analysed, 21 of which are detected in [CII]. For several of the latter the [CII] emission breaks into multiple components. For our purposes, individual clumps provide a more fair and homogeneous comparison with the spatially resolved local data and the model.  Interestingly, \citet{carniani2018} find that early galaxies are characterised by a [CII] surface brightness generally much lower than expected from the local relation, eq. \ref{dLfit}. It is worth stressing, though, that these early systems have also larger \S* values than those in the local sample.   

Model predictions, obtained from eq. \ref{eq_sigma_cii}, and using the emission model in eq.s \ref{Fcii_ionbound} and \ref{FC2D}, are presented in Fig. \ref{fig:SigmaCIISigmaSFR} along with local and high-$z$ data. We show curves for three different values of the metallicity\footnote{Recall that $Z={\cal D}=1$ indicates solar/galactic values of these quantities.}, and two values of the burstiness parameter, $k_s=1$ (panels a-c), and $k_s=5$ (panel d). The points along the curves are colour-coded to show the variation of $U$ (panel a), $N_0$ (b), and $\Sigma_g$ (c-d) as a function of \S* and \Scii. We have fixed the density of the emitting material to $n=500\, \rm cm^{-3}$ so that the local data are well matched by curves with $Z<0.5$, in agreement with the values derived by \citet{delooze2014}, i.e. $0.05 < Z < 0.4$. For reference, we also plot with a thin dashed line the curve for $Z=0.1$ and density equal to the critical density of the transition, $n=n_{\rm crit}=3027 \,\cc$. These plots condensate the core results of the present study. 

The overall shape of the predicted \Scii-\S* relation can be understood as follows. Consider moving on the curves from right (high \S* and $N_0$; low $U$) to left (low \S* and $N_0$; high $U$). Initially, \Scii\, is approximately constant as $N_F < N_0$: eq. \ref{hiZapprox} states that in this regime the [CII] flux is independent of $Z$, and only weakly increasing with $U$. As $N_0$ drops below $N_F$, at a \S* value marked by the position of the kink in the curves, \Scii\,  linearly decreases with $N_0$ (eq. \ref{lowZapprox}) at fixed $Z$. The position of the kink depends on metallicity as $N_F \propto N_d \propto Z^{-1}$ (eq. \ref{NF}), and therefore occurs at higher $N_0$ values as $Z$ decreases. Finally, the curves steepen considerably for very low \S* ($\simlt 10^{-2}  M_\odot~{\rm yr}^{-1}~{\rm kpc}^{-2}$) as the ionised layer column density, $N_i$, becomes comparable to $N_0$, thus decreasing the [CII] flux (eq. \ref{Fcii_ionbound}). When $N_0 < N_i$ the galaxy ISM is fully ionised and the \HII region is density-bounded (magenta area in the bottom-left corner). 

Local galaxies (small points) shown in Fig. \ref{fig:SigmaCIISigmaSFR}, panels (a)-(c), are located on top of our predictions for $k_s = 1$, corresponding to the KS relation. As already noted when discussing eq. \ref{Udef1}, $\log U$ decreases from $-1.6$ to $-3.0$ when \S* increases in the considered range. Stated differently, $U$ and \S* are anti-correlated as a result of the super-linear slope of the KS relation ($m=1.4)$.  By combining this information with panel (b), showing the individual column density, $N_0$, of the theoretical points, we can conclude that galaxies on the De Looze relation have gas column densities $10^{21} {\rm cm^{-2}}< N_0 < 10^{22}{\rm cm^{-2}}\, \rm cm^{-2}$. Systems with lower column densities are also characterised by larger ionisation parameters. For example, local galaxies with $Z=0.1$ and \S*$= 10^{-2}$ $ M_\odot~{\rm yr}^{-1}~{\rm kpc}^{-2}$, have $\log U \approx -2$ and $N_0 \approx 3\times  10^{21}\, \rm cm^{-2}$.

High redshift galaxies (large circles with errors) instead populate the high \S* region of the plot. They tend to have lower ionisation parameters ($\log U < -2.5$), and large gas surface/column gas densities (see panels b-c), with $N_0 \approx 10^{22.5}{\rm cm^{-2}}$. So it appears that observed early galaxies have noticeably different structural properties compared to local ones. We return to this point in Sec. \ref{under}.  

We can also identify the region of the parameter space occupied by density-bounded systems in which the gas is fully ionised. Using the relations provided above (e.g. eq. \ref{FC2D}), we then impose the condition $N_i > N_0$ shown in Fig. \ref{fig:SigmaCIISigmaSFR} as a magenta shaded area.
The density-bounded area is relatively small, and it includes systems with low \S* (corresponding to gas column densities $< 10^{21} \rm cm^{-2}$) and high $U$ values. Hence, from the curves in panels (a)-(c) for which $\ks = 1$, it appears that none of the galaxies, both in the local or high-$z$ sample, is density-bounded. Density-bounded systems tend to have a low \CII surface brightness as the line can only originate in the ionised layer (Zone I) where carbon is largely in the form of \CIIIion. 

As already pointed out, in general the contribution from ionized gas to the total \Scii is negligible. However, as we consider galaxies closer or into to the density-bounded regime, such contribution increases to reach 100\%; this occurs for log \S* $< -2$. Interestingly, this result is in line with the conclusions reached by an increasing number of studies who find that in a sample of nearby galaxies the fraction of [CII] emission arising from the ionized gas varies from $<10$\% in systems with log \S* $> -1$ to 20-30\% in galaxies with log \S* $< -2.5$  \citep{Diaz2017, Croxall17, Parkin13, Hughes15}.

Finally, let us analyse the effects of increasing $\ks$. In panel (d) we set $\ks=5$. In this case a galaxy with a given $\Sigma_g$ has a star formation rate that is 5$\times$ larger than expected from the KS relation, a situation resembling a starburst galaxy. An increase of $\ks$ produces a rightwards shift of all the curves at various metallicities. Perhaps, a more meaningful way to interpret the effect is that at a given \S* a galaxy has a lower \CII luminosity as a result of the paucity of gas, and of the more extended \HII layer. Hence, the galaxy drops considerably below the De Looze relation.
Finally, note that for $\ks=5$, the density-bounded limit shifts to higher \S*.

\subsection{Why are high-$z$ galaxies [CII]-underluminous?}\label{under}
We now turn to a more specific comparison with the data and then concentrate on the deviation of high-$z$ sources from the De Looze relation. The local data  are well fit by the curve with metallicity (or dust-to-gas ratio) $Z = 0.1$, which is consistent with the mean value $0.05 \simlt Z \simlt 0.4$ deduced by \citet{delooze2014}. However, note that the theoretical curves and the data show a curvature/shape that is only marginally caught by the power-law fit adopted by those authors (thin line in Fig. \ref{fig:SigmaCIISigmaSFR}). 

In general, an increasing trend of specific \CII luminosity with $Z$ is seen for low/moderate values of \S*. For larger surface star formation rates \Scii\, becomes independent of $Z$, and saturates at $\approx 10^7 L_\odot {\rm kpc}^{-2}$. Apart from the weak logarithmic dependence on $U$, the saturation value depends only on density; it can be estimated by combining eq. \ref{hiZapprox}  and \ref{eq_sigma_cii}: 
\be
\Sigma_{\rm [CII]} \approx 1.8\times 10^7 \left(\frac{n}{500\, \cc}\right)
\ln\left(\frac{U}{10^{-3} }\right) \quad L_\odot {\rm kpc}^{-2}.
\label{sat}
\ee
By setting $n=n_{\rm crit}=3027\, \cc$ we find that the surface brightness reaches \Sciimax$\approx 10^8 L_\odot {\rm kpc}^{-2}$. 

Thus, the model predicts that at high \S* the data should deviate from the power-law trend expected from the empirical fit. Indeed, there is a hint from the local data that this might be the case. However, we consider this agreement only as tentative given that in that regime the \citet{delooze2014} sample  contains only one galaxy (NGC 1569, rightmost data points at log \S* $\approx 0$). 

\begin{figure}
    \centering
    \includegraphics[scale=0.21]{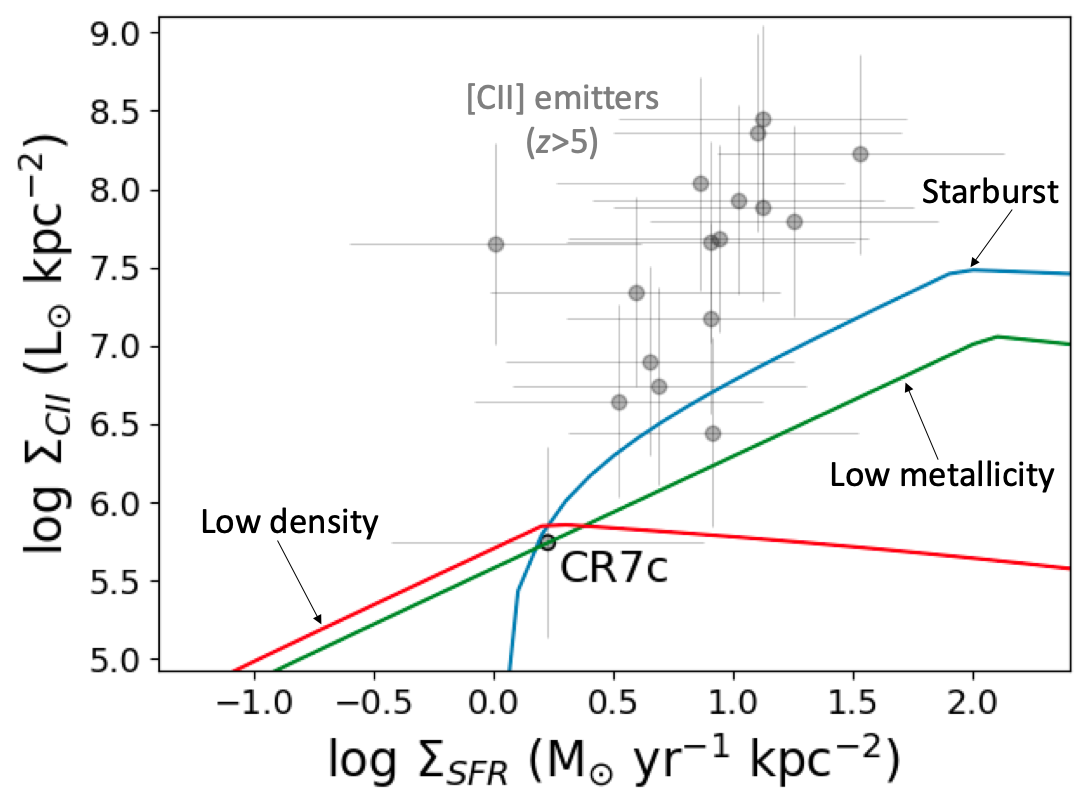}
    \caption{Degeneracy of the observed \Scii\, with the three free parameters, $n$, $Z$ and $k_s$. In this example, the data point corresponding to CR7c ($z=6.6$) can be equally well interpreted by three different models assuming (a) low ($n=20\,\cc$) density, (b) low metallicity, $Z=0.003$, or (c) strong starburst ($k_s=25$). See text for more details.}
    \label{degeneracy}
\end{figure}

The saturation effect is crucial to interpret $z>5$ galaxies. These early systems have large surface star formation rates, \S*$> 1 M_\odot {\rm yr}^{-1} {\rm kpc}^{-2}$. Then, our model predicts that, if their metallicity is $Z\simgt 0.1$, they are in the saturated emission regime governed by eq. \ref{sat}. The saturation is also one of the factors (see below) that can explain why high-$z$ galaxies lie below the local De Looze relation (thin dashed curve in Fig. \ref{fig:SigmaCIISigmaSFR} with a measured \Scii $<$ \Sciimax. Following this interpretation, the mean density of the emitting gas in these early galaxies should be in the range $20\, {\cc} < n \simlt n_{\rm crit}$, with the most diffuse system being CR7c ($z=6.6$), according to \citet{carniani2018} classification, whose \Scii$\approx 10^6 L_\odot {\rm kpc}^{-2}$. Fig. \ref{degeneracy} (red curve) shows the result of the model fit to the CR7c point, which is essentially independent of metallicity.  As a consequence of the low gas density, the emitting regions in CR7c have a large size, $\ell_0 \approx 100$ pc. In addition, CMB suppression effects of the line emission from such a low density gas can become important \citep{kohandel:2019}. 

However, explaining low \Scii\, systems as due to a low gas density is not the only option.
From [CII] observations alone, it cannot be excluded that the faintest systems can instead have an extremely low metallicity. In this case they could be fit by model curves in which \Scii\, is still raising with metallicity.
This is illustrated by the green curve in Fig. \ref{degeneracy}; the corresponding model parameters are $n=500\,\cc, k_s=1$ and a metallicity $Z=0.003$, a value close to the metallicity of the intergalactic medium at $z=6$ \citep{Dodorico13}. Additionally, simulations of high-$z$ galaxies \citep{pallottini:2019} shows that the presence of low metallicity regions plays a sub-dominant role in determining the galaxy deviation from the De Looze relation.
Hence, we consider this explanation unlikely, albeit we point out that at least some degeneracy exists between density and metallicity in determining the observed [CII] surface brightness. Yet a third option is possible, i.e. a larger $k_s$ corresponding to a starbursting phase. This case is represented by the blue curve, with the fiducial values $n=500\,\cc, Z=0.1$ but a $k_s=25$. 

Fig. \ref{degeneracy} is only meant as an illustration of the degeneracy existing among $n, k_s$ and $Z$. We have used the most extreme system (CR7c) to show how different combinations of these parameters might produce the observed low \Scii\, value. Clearly all the other high-$z$ systems might be interpreted in the same way. 
In summary, our model predicts that the evidence that high-$z$ galaxies have lower-than-expected [CII] surface brightness can be explained by a combination of these three factors: (a) large upward deviations ($\ks \gg 1$) from the KS relation, implying that they are in a starburst phase; (b) low metallicity; (c) low gas density, at least for the most extreme sources (e.g. CR7). 
These results support the similar suggestions made by \cite{vallini2015}. Large $\ks$ values might result from negative stellar feedback effects evacuating the gas around star formation sites and/or from enhanced periods of star formation activity following merger events.

\section{Summary}

We have developed an analytical model to interpret \CII line emission which can be applied both to low and high redshift galaxies. First, we have characterised the ionisation/PDR structure properties and determined the extent of three physically distinct (ionised, neutral/molecular, dark) regions in terms of the metallicity, $Z$ (or dust-to-gas ratio, $\cal D$), and ionisation parameter, $U$, of the gas. Then we predict the \CII line emission for both ionisation- and density-bounded conditions. We have successfully validated the model against numerical radiative transfer calculations performed with \code{cloudy}.

Once properly cast in terms of observed quantities, such as \CII surface brightness (\Scii), and star formation rate per unit area (\S*), the model has been used to interpret the observed \Scii -- \S* relation, and the deviations from it observed for galaxies in the EoR. We find that:
\begin{itemize}
  \item There is an overall increase in \Scii~with $Z$ and \S*. However, for \S*$\simgt 1 M_\odot~{\rm yr}^{-1}~{\rm kpc}^{-2}$, \Scii~saturates (see Fig. \ref{fig:SigmaCIISigmaSFR}) at a level $\approx 10^7 L_\odot {\rm kpc}^{-2}$, which depends linearly on the density of the emitting gas (fiducially assumed to be $n=500\, \rm cm^{-3}$ to match the $z\approx 0$ dwarf sample). As a result, the relation has a more complex shape than the simple power-law usually assumed to fit the data.
  \item The \Scii -- \S* relation can be read as a sequence of decreasing $U$ with increasing \S*. Galaxies with $\Sigma_{\rm SFR}< 10^{-2.5} M_\odot~{\rm yr}^{-1}~{\rm kpc}^{-2} $ are predicted to be highly ionised due to their low gas column densities.
  \item Upward deviations from the KS relation, parametrized by the \quotes{burstiness} parameter $\ks$ (eq. \ref{KS}), shift the predicted \Scii~ towards higher \S*, causing galaxies at a fixed \S* to have unexpectedly low [CII] surface brightness.
  \item Our model predicts that under-luminous \CII systems, as those routinely observed at high-$z$, can result from a combination of these three factors: (a) large upward deviations ($\ks \gg 1$) from the KS relation, implying that they are in a starburst phase; (b) low metallicity; (c) low gas density, at least for the most extreme sources (e.g. CR7).     
\end{itemize}
Observations of \CII emission alone cannot break the degeneracy among the above three parameters, although extreme deviations from the De Looze relation might imply unrealistic conditions based on additional considerations. Hence to fully characterise the properties of the interstellar medium of galaxies additional and complementary information must be sought.
This can be obtained, for example, by combining \CII observations with other FIR lines, like e.g. [OIII] \citep{inoue2016,vallini2017}, dust continuum (\citealt{behrens2018}, to constrain $\cal D$), CO lines \citep{vallini2018}, or even H$_2$ lines, hopefully becoming available at these redshifts with SPICA \citep{spinoglio:2017,Egami18}. An interesting, and more readily available alternative are optical nebular lines, such as Ly$\alpha$ and \CIII, used in combination with machine learning strategies \citep{Ucci19}. The implications of these measurements still need to be worked out fully; we defer this study to future work.

\section*{Acknowledgements}
AF and SC acknowledge support from the ERC Advanced Grant INTERSTELLAR H2020/740120. LV acknowledges funding from the European Union's Horizon 2020 research and innovation program under the Marie Sk\l{}odowska-Curie Grant agreement No. 746119. 
This research was supported by the Munich Institute for Astro- and Particle Physics (MIAPP) of the DFG cluster of excellence \quotes{Origin and Structure of the Universe}.



\bibliographystyle{mnras}
\bibliography{Revised} 

\appendix
\section{Carbon ionisation}\label{appendix:c_ionization}

We derive the fraction of singly ionised carbon $x_{\rm C}\equiv n_{\rm CII}/n_{\rm C}$ as a function of the neutral fraction, $\fhi\equiv n_{\rm HI}/n$, in the \HII region. 
Consider the carbon ionisation equilibrium equation: 
\begin{equation}
    \Gamma_{\rm C} n_{\rm CII} = \alpha_{\rm C} n_{\rm CII} n_{\rm CIII}
\label{Cionization}
\end{equation}
where we account for the $\CIIion \Leftrightarrow \CIIIion$ equilibrium only. In eq. \ref{Cionization}, $\Gamma_{\rm C}$ is the optically-thin carbon photo-ionisation rate:
\be
\Gamma_{\rm C}= \int_{\nu_1}^\infty   \frac{F_\nu}{ h_P \nu} \sigma^{\rm C}_\nu d\nu \approx \frac{F_L}{h_P} \nu_{1} \sigma_{\rm C}=Unc\sigma_{\rm C}
\label{Gamma1}
\ee
where the CI photo-ionisation cross-section at $h\nu_1=11.26$ eV is $\sigma_{\rm C} =  3.7 \times 10^{-18}\rm \, cm^{-2}$ \citep{spitzer1978book}, and $\alpha_{\rm C} = 6.02 \times 10^{12}\, \rm cm^{3} \, s^{-1}$ is the recombination coefficient \citep[][Tab. 5]{nahar1997}. Analogously, write $\Gamma=Unc\sigma_{\rm H}$.
From eq. \ref{Cionization}, assuming $n_e\approx n_p$ we derive:
\be
n_e = \frac{\Gamma_{\rm C} x_{\rm C}}{\alpha_{\rm C} (1-x_{\rm C})}
\ee
which, once substituted in the photo-ionisation equilibrium equation $\Gamma n = \alpha_B n_e n_p$ (see eq. \ref{nHI}) yields:
\begin{equation}
    \Gamma n = \alpha \left[ \frac{\Gamma_{\rm C} x_{\rm C}}{\alpha_{\rm C} (1-x_{\rm C})}\right]
\end{equation}
We now need to invert this equation to get $x_c(\fhi)$:
\begin{equation}
    \left(\frac{\Gamma n}{\alpha}\right)^{1/2} \fhi^{1/2} = \frac{\Gamma_{\rm C} x_{\rm C}}{\alpha_{\rm C} (1-x_{\rm C})}
\end{equation}
which can be rewritten as
\begin{equation}
    \left(\frac{\alpha_{\rm C}}{\Gamma_{\rm C}}\right) \left(\frac{\Gamma }{\alpha}\right)^{1/2} n^{1/2} \fhi^{1/2} = \frac{x_{\rm C}}{1-x_{\rm C}}.
\end{equation}
Define $B\equiv  (\alpha_{\rm C}/\Gamma_{\rm C}) ({\Gamma}/{\alpha})^{1/2} = 0.046/\sqrt{Un}$; we finally get
\begin{equation}
    x_{\rm C} = \left[\frac{B n^{1/2}}{1+ B n^{1/2} \fhi^{1/2}}\right] \fhi^{1/2} = \frac{\zeta}{1+\zeta}
\end{equation}
where $\zeta=0.046 (\fhi/U)^{1/2}.$

\section{Line collisional excitation}\label{appendix:cooling_functions}

We derive here the emission from the illuminated slab in two collisionally-excited ({\it forbidden } or {\it semi-forbidden}) lines of \CII 158 $\mu$m and \CIII 1909 \AA. The collisional de-excitation cross-section from the upper (2) to lower (1) state, $\sigma_{21} \propto v^{-2}$ of the electrons due to Coulomb focusing:
\be
\sigma_{21} = \frac{\pi h^2}{m_e^2 v^2} \frac{\Omega(1,2; E)}{g_1}, 
\ee
where $\Omega(1,2; E)$ is the \textit{collision strength} symmetrical in (1,2). It must be computed quantum-mechanically, and is approximately constant close to the resonance energy, $E_{12}$. The rate of de-excitations ($\cc s^{-1}$) is written as
\be
R_{21} = n_e n_2 \int_0^\infty \sigma_{21} v(E) f(E)dE \equiv n_e n_2 q_{21},
\ee
where 
\be
f(E) = \frac{2E^{1/2}}{\pi^{1/2}(kT)^{3/2}} e^{-E_{12}/kT}
\ee
is the Maxwellian distribution. By substituting the expression for the cross-section and $f(E)$, and recalling that $v=\sqrt{2E/m}$  we find
\be
q_{21} = \frac{\beta}{ \sqrt{T}}\frac{\Upsilon}{ g_2},
\ee
where 
$\beta = (2\pi \hslash^4/m_e^3 k)^{1/2} = 8.629 \times 10^{-6}$, and
\be
\Upsilon(T) = \int_0^\infty \Omega_{21}(E) e^{-E/kT} d\left(\frac{E_{12}}{ kT}\right),
\ee
The collisional excitation coefficient in thermodynamic equilibrium can be simply derived from the de-excitation one:
\be
q_{12} = \frac{g_2}{ g_1} q_{21} e^{-E_{12}/kT} =  \frac{\beta}{ \sqrt{T}}\frac{\Upsilon}{ g_1}  e^{-E_{12}/kT}.
\ee
The detailed balance equation of the levels population reads:
\be
n_e n_1 q_{12} = n_e n_2 q_{21} +A_{21} n_2.
\ee
For densities below the critical one, $n_{\rm crit}=A_{21}/q_{21}$ we can write the following approximation:
\be
\frac{n_2}{ n_1} = \frac{n_e q_{12}}{ A_{21} + n_e q_{21}} \approx n_e \frac{q_{12}}{ A_{21}}  
\ee
The power emitted in the line per unit volume is
\be
L = n_2 A_{21} E_{12} = n_e n_1 q_{12} E_{12} \equiv n_e n_1 \Lambda(T),
\ee
i.e. for densities below the critical density every collisional ionisation results in a photon emission. The full expression for the cooling function (erg $\rm cm^3 \rm s^{-1}$) is
\be
\Lambda(T) =  \frac{\beta}{ \sqrt{T}}\frac{\Upsilon(T)}{ g_1} E_{12} e^{-E_{12}/kT}.
\label{lambda}
\ee
Specialize to the emission from [CII] and CIII] lines. The first is the $^2P_{3/2} \rightarrow\, ^2P_{1/2}$ \CII 157.78 $\mu$m transition for which the statistical weights $(2J+1)$ are $g_2=4$ and $g_1=2$; the second is the \CIII 1908.7 \AA\, $^1P_{1} \rightarrow\, ^1S_{0}$ transition, whose statistical weights are $g_2=3$ and $g_1=1$. The energy separation for \CII is $T_*=91.92$ K ($E_{12}=0.0079$ eV), while for \CIII it is $T_*=7.59\times 10^4$K ($E_{12}=6.54$ eV).

For the maxwellian-averaged collision rates, $\Upsilon$, we use the following expressions, taken from \citet{goldsmith12} and \citet{osterbrock1992} for the two ions:
\[\begin{array}{lcr}\label{cr}
\Upsilon_{\rm [CII]}^e(T) & = &  0.67 T^{0.13}\\
\Upsilon_{\rm [CII]}^H(T) &= & 1.84\times 10^{-4} T^{0.64}\\
\Upsilon_{\rm CIII]}^e(T) & = &1.265-0.0174 \times 10^{-4} T \end{array}\]

\begin{figure}
    \centering
    \includegraphics[scale=0.38]{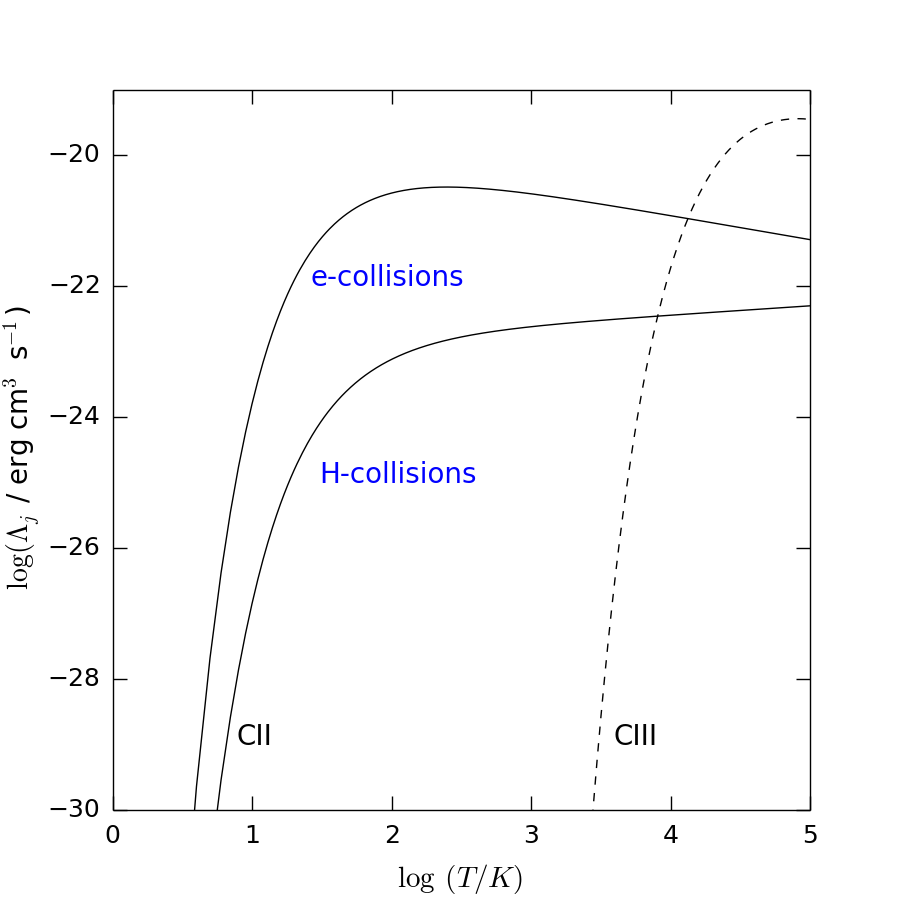}
    \caption{\CII and CIII] cooling function vs. gas temperature for excitations due to collisions with electrons. For \CIIion we also show the analog quantity for collisions with H atoms as indicated be the labels. }
    \label{fig:cool}
\end{figure}

In the present study, unless otherwise specified, we will use the values of $T=10^2$ K ($10^4$ K) for \CIIion (\CIIIion) as this ion is predominantly located in neutral (ionized) regions (see Fig. \ref{fig:temp_profiles}). Hence, from eq. \ref{lambda} we obtain, $\Lambda_{\rm [CII]}^{(2)} = \Lambda_{\rm [CII]}(T=10^2 \mathrm{K}) = 7.65 \times 10^{-24}$ erg cm$^3 \rm s^{-1}$, and $\Lambda_{\rm CIII]}^{(4)}=\Lambda_{\rm [CIII]}(T=10^4 \mathrm{K}) = 1.88 \times 10^{-22}$ erg cm$^3 \rm s^{-1}$. Note that for \CII we have considered excitations by collision with H atoms. In \HII regions, the \CII line is excited by electron collisions; therefore, $\Lambda_{\rm [CII]}^{(4)}= \Lambda_{\rm [CII]}(T=10^4 \mathrm{K}) = 1.2 \times 10^{-21}$ erg cm$^3 \rm s^{-1}$. The temperature dependence of the above cooling functions is shown in Fig. \ref{fig:cool}. 
%

\bsp	
\label{lastpage}
\end{document}